\documentclass[useAMS,usenatbib]{mn2e}

\usepackage{amssymb}
\usepackage{amsfonts}
\usepackage{natbib}
\usepackage{epsfig}
\usepackage{mncite}
\usepackage[varg]{txfonts}

\newcommand{\de}{\mbox{d}}
\newcommand{\msun}{M_{\sun}}

\begin{document}

\title[Black hole mergers: the final parsec problem]{Black hole
  mergers: can gas discs solve the `final parsec' problem?}

\author[Lodato et al] {G. Lodato$^{1,2}$, S. Nayakshin$^1$,
  A.R. King$^1$ \& J. E. Pringle$^{1,3}$\\ $^1$Department of Physics
  \& Astronomy, University of Leicester, Leicester, LE1 7RH, UK\\ $^2$
  Dipartimento di Fisica, Universit\`a di Milano, Via Celoria, 16,
  Milano, I-20133, Italy\\ $^3$ Institute of Astronomy, Madingley
  Road, Cambridge, CB1 0HA, UK}

\maketitle

\begin{abstract}

  We compute the effect of an orbiting gas disc in promoting the
  coalescence of a central supermassive black hole binary. Unlike
  earlier studies, we consider a finite mass of gas with explicit time
  dependence: we do not assume that the gas necessarily adopts a
  steady state or a spatially constant accretion rate, i.e. that the
  merging black hole was somehow inserted into a pre--existing
  accretion disc.  We consider the tidal torque of the binary on the
  disc, and the binary's gravitational radiation. We study the effects
  of star formation in the gas disc in a simple energy feedback
  framework.
  
  The disc spectrum differs in detail from that found before. In
  particular, tidal torques from the secondary black hole heat the
  edges of the gap, creating bright rims around the secondary. These
  rims do not in practice have uniform brightness either in azimuth or
  time, but can on average account for as much as 50\% of the
  integrated light from the disc. This may lead to detectable
  high--photon--energy variability on the relatively long orbital
  timescale of the secondary black hole, and thus offer a prospective
  signature of a coalescing black hole binary.

  We also find that the disc can drive the binary to merger on a
  reasonable timescale only if its mass is at least comparable with
  that of the secondary black hole, and if the initial binary
  separation is relatively small, i.e. $a_0 \lesssim 0.05$ pc. Star
  formation complicates the merger further by removing mass from
  the disc. In the feedback model we consider, this sets an effective
  limit to the disc mass. As a result, binary merging is unlikely
  unless the black hole mass ratio is $\la 0.001$. Gas discs thus
  appear not to be an effective solution to the `last parsec' problem
  for a significant class of mergers.

\end{abstract}
\begin{keywords}
  accretion, accretion discs -- black hole physics -- galaxies:
  formation -- cosmology: theory -- instabilities -- hydrodynamics
\end{keywords}

\section{Introduction}

In recent years, the process of shrinking a supermassive black hole (SMBH)
binary by interaction with a circumbinary gaseous disc has been the topic of
intense theoretical research. Initially this reflected attempts to overcome
the `last parsec' problem, i.e. the fact that dynamical friction with the
stellar background is ineffective in shrinking the binary below separations
smaller than 1pc \citep{begelman80,milos01}, to the point where gravitational
radiation can complete the coalescence of the two holes. More recently there
has been interest in finding an electromagnetic counterpart to the
gravitational wave emission expected during the final stages of black hole
coalescence
\citep{ivanov99,natarajan02,milosavljevic05,dotti06,loeb07,CuadraEtal09}.

In general most work on this subject has assumed driving by an
accretion disc with constant mass inflow rate supplied from distances
far from the binary, and thus effectively assumed an infinite mass
supply. However it is clear that in reality, where the gas is
part of the galaxy merger producing the SMBH binary, the mass of gas
ending up in the disc must be finite. Thus in general the disc does not
settle to a steady state with a constant accretion rate, and its
structure differs from a standard disc because it is affected by the
tidal torque exerted by the binary. Moreover the rate or even the
success of the shrinkage must depend on the total disc mass (as hinted
by Cuadra et al, 2009). Here we investigate the problem by taking a
disc of finite mass which is explicitly time--varying. We find
significant differences from the results of assuming a steady--state
accretion disc.

We model the disc evolution in terms of a simple diffusion equation,
including the effects of the tidal torques from the secondary and
calculating the binary orbit evolution self--consistently. Our initial
condition is a finite disc mass concentrated at radii of the order of
the initial binary separation. While still highly idealised, we regard
this choice as more realistic than the assumption of a steady state,
constant--$\dot{M}$ disc, into which a second black hole has somehow
been inserted.

The paper is organised as follows. In Section 2 we describe the
general features of binary shrinkage by a gas disc. In Section 3 we
describe our method for following both the binary and disc evolution.
In Section 4 we describe the results. We constrain the disc mass
required to bring the binary to coalescence and describe the expected
appearance of such discs. In Section 5 we draw our conclusions.

\section{Disc assisted binary shrinkage}

The process of disc assisted binary shrinkage (`hardening') is for low
mass ratios dynamically similar to the process of planetary migration
within a protostellar disc, and many studies have alluded to this
analogy. For a low (but finite) mass companion, the tidal
interaction between the binary and the gaseous disc produces a gap in
the disc at the location of the secondary. The exchange of angular
momentum across this gap is mediated through tides by the `satellite'
(i.e. the secondary black hole) which behaves as a fluid element in
the disc. In this case, the binary shrinks on the disc's viscous
timescale $t_{\nu}$ \citep{natarajan02}. For a $10^8M_{\odot}$ primary
black hole the viscous time at $R\sim 1$ pc is a few times $10^8$
years, so the binary can shrink in a reasonable time before
gravitational radiation emission takes over as the main cause of
shrinkage.

However things are different if the angular momentum associated with
the binary orbit is comparable to or larger than the local disc
angular momentum (for a circular binary orbit, this is equivalent to
requiring the secondary mass to be comparable to or larger than the
local disc mass). Then the viscous torques in the disc are not
able to redistribute efficiently the excess angular momentum
transferred by the tides to the disc, and the binary shrinking slows
down significantly. 

We define $M_{\rm p}$ and $M_{\rm s}$ as the masses of the primary and
secondary black holes. In general the binary orbit evolution occurs on
a timescale \citep{clarkesyer95,ivanov99}:
\begin{equation}
t_{\rm shrink}=\frac{M_{\rm d}(a)+M_{\rm s}}{M_{\rm d}(a)}t_{\nu},
\label{eq:shrink}
\end{equation}
which for $M_{\rm s}\gg M_{\rm d}(a)$ can be significantly longer than the
disc viscous time. In the equation above the local `disc mass' is defined as
$M_{\rm d}(a)=4\pi\Sigma(a) a^2$, where $\Sigma(a)$ is the surface density of
the unperturbed disc at the binary separation $a$. In such a configuration,
disc--assisted shrinking from parsec scales would take much longer than $\sim
10^9$ years and would effectively preclude binary coalescence. Here the
companion acts as a dam in the accretion flow. The disc inside the binary
orbit is rapidly accreted, since it cannot be refilled from the outer
circumbinary disc. Outside the binary orbit, accreting gas builds up at the
circumbinary disc's inner edge. This moves inwards on a timescale $\sim t_{\rm
  shrink}$, much longer than the viscous time.  

If we were now additionally to make the usual assumption that the disc is
continuously fuelled from the outside, then the surface density at the inner
edge of the circumbinary disc would increase until it was large enough to
speed up the binary hardening, as described by \citet{ivanov99}. The disc
structure would be such that $\nu\Sigma = \dot{M}/2\pi = {const.}$ at large
radii, while at smaller radii $\nu\Sigma\propto R^{-1/2}$, typical of a
`decretion' disc \citep{pringle91}, where $\nu$ is the disc viscosity.
However, in reality, with a finite gas mass, the surface density at the inner
edge of the circumbinary disc is unable to grow steadily at the rate described
by \citet{ivanov99}. Evolution proceeds more slowly and a numerical solution
is needed. We discuss these solutions in Section 3 below. The overall
normalisation of the surface density differs from that predicted by
\citet{ivanov99}, while the scaling with radius does follow the
$\nu\Sigma\propto R^{-1/2}$ decretion--like solution.

The qualitative appearance of the system as the binary evolves towards
coalescence is clear. In a major merger two black holes of comparable
mass are brought together at a distance $\lesssim 0.1$pc
\citep{escala05}, together with a finite amount of gas which we assume
to settle rapidly to a fairly wide disc configuration. Part of the gas
can accrete directly on to the primary black hole, producing a burst
of AGN--like activity. Given the large amount of gas mass available,
this burst is likely to be Eddington limited. As a result the disc
around the primary black hole may be dispersed by the strong outflows
arising from such a super-Eddington flow. Since the secondary hole
acts as a dam for the gas outside the binary, the disc around the
primary is not readily refilled. For protostellar accretion, the
extent to which the dam is porous is still under discussion
\citep[e.g.,][]{LDA06}. Since AGN discs are typically much thinner
than protostellar discs, the porosity of the dam for the AGN discs
considered here is likely to be much reduced, and we neglect this
effect for the purposes of this paper. Thus after this initial
episode, the system is characterised by a massive circumbinary disc
truncated at the binary separation, and possibly by a low density
circumprimary disc left over from the initial Eddington-limited
episode. The system appears bright at long wavelengths (optical/IR),
qualitatively similar to the description of \citet{milosavljevic05}.

At this point there are two possibilities: either (a) the
circumbinary disc mass is too low, the coalescence fails (or takes
inordinately long), and the two black holes are left orbiting at a
fraction of a parsec distance, or (b) the disc mass is large enough
to drive the binary to coalescence on a reasonable timescale. In
this event the peak of the disc spectral energy distribution
progressively moves to higher energies and the disc becomes more
luminous, until gravitational radiation losses bring the black holes
into coalescence. If the remnant circumprimary disc is still
present, it will suddenly brighten as it is squeezed by the merging
binary, as qualitatively described by \citet{natarajan02}. After
coalescence the dam on the accretion flow is removed, so the inner
disc hole is refilled and the disc spectrum finally moves to high
energies.

\section{Time--dependent model}

\subsection{Coupled disc/SMBH binary evolution}

The evolution of an accretion disc in the presence of an embedded
satellite can be described by a diffusive evolution model for the
disc, including the tidal term arising from the secondary, i.e.
\begin{equation}
\label{eq:diffplanet}
  \frac{\partial\Sigma}{\partial
  t}=\frac{3}{R}\frac{\partial}{\partial R}
  \left(R^{1/2}\frac{\partial}{\partial R}(R^{1/2}\nu\Sigma)\right)
  -\frac{1}{R}\frac{\partial}{\partial
  R}\left(2\Omega R^2\lambda\Sigma\right),
\end{equation}
where $\lambda=\Lambda/(\Omega R)^2$, with $\Lambda$ the specific
tidal torque, and
\begin{eqnarray}
\label{eq:torque}
\lambda= & \displaystyle\frac{q^2}{2}\left(\displaystyle
\frac{a}{p}\right)^4  &
R>a \\
\nonumber \lambda= & -\displaystyle\frac{q^2}{2}\left(\displaystyle 
\frac{R}{p}\right)^4  & R<a. 
\end{eqnarray}
(We choose this formulation rather than the standard one with
$\Lambda$, for later convenience.) In equation (\ref{eq:torque}),
$\Omega\approx \sqrt{GM_{\rm p}/R^3}$ is the angular velocity at
radius $R$, $a$ is the radial position of the satellite,
$q=M_{\mathrm{s}}/M_{\rm p}$ is the mass ratio between the secondary
and the primary black hole, and $p=R-a$. This simplified form of the
specific torque is commonly used in literature
(see, e.g.,
\citealt{armitage2002,armibonnel2002,natarajan02,LC04}). We smooth the
torque term for $R\approx a$, where it would have a singularity (see
equation \ref{eq:torque}). We use the same smoothing prescription as
in \citet{clarkesyer95} and \citet{linpap86b}, i.e. for
$|R-a|<\max[H,R_{\mathrm{H}}]$, where $H$ is the disc thickness and
$R_{\mathrm{H}}$ is the size of the Hill sphere (Roche lobe) of the
secondary. Finally we note that if the mass ratio $q$ is large, the
formalism above implies a non-vanishing torque at relatively large
distances from the secondary. However, the tidal torque is actually
provided by the cumulative effect of a series of Lindblad resonances
\citep{goldreich80}, and should therefore vanish beyond the outer and
the inner Lindblad resonance, located at $p=2^{\pm2/3}a$. We therefore
smoothly truncate the torque beyond these radii.

The back reaction of the disc on the secondary orbital motion follows
from angular momentum conservation, in the form
\begin{eqnarray}
\label{eq:planet}
\frac{\de}{\de t}(M_{\mathrm{s}}\Omega_{\mathrm{s}}a^2) & = &
-\int_{R_{\mathrm{in}}}^{R_{\mathrm{out}}}2\pi \Omega^2 R^3\lambda\Sigma \de R
\\ \nonumber & = & -2\pi GM_{\rm p} \int_{R_{\rm in}}^{R_{\rm
    out}}\lambda\Sigma\mbox{d}R,
\end{eqnarray}
where the integral is taken over the whole disc surface and
$\Omega_{\rm s}$ is the angular velocity of the secondary. It follows
that the behaviour of the system described here is determined by two
dimensionless parameters. First,
\begin{equation}
A=\frac{\Omega R^2}{\nu}q^2,
\end{equation}
measures the relative strength of the second and first terms on the
rhs of equation (\ref{eq:diffplanet}), while
\begin{equation}
B=\frac{4\pi\Sigma(a) a^2}{M_{\mathrm{s}}}
\end{equation}
gives a measure of the magnitude of the rhs of equation
(\ref{eq:planet}).

The parameter $A$ has the simple physical interpretation
\citep{linpap79a} $A=(\Delta/R)^3$, where $\Delta$ is the gap
width. To open a gap, the gravitational effect of the satellite has to
overcome the pressure of the disc and the viscosity, which both oppose
gap opening. We neglect the effects of pressure in this paper. 

The second important parameter, $B$, represents the ratio of local
disc mass to satellite mass. If the satellite is more massive than the
disc (i.e. $B\ll 1$) its inertia makes migration slow. Conversely, if
$B\gg 1$, the secondary behaves like a fluid element of the disc
and migrates on the local viscous timescale.

\subsection{Gravitational wave torques}

We also include in our calculation the torques arising from
gravitational wave radiation. For this purpose we add in
Eq. (\ref{eq:planet}) the gravitational wave term causing the binary
separation to evolve at a rate
\begin{equation}
  \left. \frac{\mbox{d}a}{\mbox{d}t} \right|_{\rm gw} 
  = - \frac{64}{5}\frac{G^3M_{\rm
      p}^3}{c^5 a^3}\frac{q}{(1+q)^2}.
\end{equation}
The corresponding timescale for a coalescence induced purely by gravitational
wave emission is thus
\begin{eqnarray}
t_{\rm gw} & = & \frac{5}{64}\frac{c^5a^4}{G^3M_{\rm
    p}^3}\frac{(1+q)^2}{q} \\ \nonumber \displaystyle & \approx &
2~10^{11}\left(\frac{a}{0.1\mbox{pc}}\right)^4\left(\frac{M_{\rm
    p}}{10^8M_{\odot}}\right)^{-3}\frac{(1+q)^2}{q} \mbox{yr}.
\end{eqnarray}

\subsection{Energy dissipation from tides}

Most previous studies of satellite--disc interaction using a formalism
like that above do not discuss in detail the extra energy dissipation
introduced by the tidal term. As a consequence, the disc spectrum in
the presence of a massive satellite is taken to be that of a standard
accretion disc truncated at a radius of the order of the binary
separation (e.g., see \citealt{milosavljevic05}). 

In reality this point is subtle because the tidal interaction is not
formally a dissipative term. However the formalism of Eqs. (2-4)
enforces angular momentum conservation, and simultaneously assumes
that the disc and secondary orbits remain circular. This implicitly
assumes the presence of some dissipative term to damp out any
eccentric motion. To compute this extra dissipation, we ignore the
viscous term in equation (\ref{eq:diffplanet}) and follow the
evolution of the system as driven purely by the tidal term. The total
energy of the disc (kinetic plus potential) is:
\begin{equation}
\label{eq:discenergy}
E_{\rm d} = -2\pi\int_{R_{\rm in}}^{R_{\rm out}}\frac{GM_{\rm p}}{2R}\Sigma
R\mbox{d}R = -\pi GM_{\rm p}\int_{R_{\rm in}}^{R_{\rm out}} \Sigma \mbox{d}R,
\end{equation}
and its time derivative is
\begin{eqnarray}
\label{eq:discenergychange}
\frac{\mbox{d} E_{\rm d}}{\mbox{d} t} & = & -\pi G M_{\rm p}\int_{R_{\rm
    in}}^{R_{\rm out}}\frac{\partial\Sigma}{\partial t} \mbox{d} R\\ 
\nonumber & = & \displaystyle \pi GM_{\rm p}\int_{R_{\rm in}}^{R_{\rm
    out}}\frac{\mbox{d}R}{R}\frac{\partial}{\partial R} \left(2\Omega
R^2\lambda\Sigma\right) \\ 
\nonumber & = & \displaystyle 2 \pi G M_{\rm
  p}\int_{R_{\rm in}}^{R_{\rm out}} \Omega \lambda\Sigma\mbox{d}R,
\end{eqnarray}
where in the last step we have integrated by parts. We now compute the
change in the energy $E_{\rm s} = -GM_{\rm p}M_{\rm s}/2a$ of the
secondary. Using Equation (\ref{eq:planet}) we have
\begin{equation}
\label{eq:secondaryenergy}
\frac{\mbox{d} E_{\rm s}}{\mbox{d} t} = \frac{GM_{\rm p}M_{\rm
    s}}{2a^2} \frac{\mbox{d}a}{\mbox{d}t} = -2\pi GM_{\rm
  p}\Omega_{\rm s} \int_{R_{\rm in}}^{R_{\rm
    out}}\lambda\Sigma\mbox{d}R.
\end{equation}
We note that the energy changes of the disc and the satellite differ
by a factor $\Omega-\Omega_{\rm s}$. Therefore the total energy loss
due to the effect of tidal torques in Equations (2-4) is given by:
\begin{equation}
\label{eq:tidaldiss}
-\frac{\mbox{d}}{\mbox{d}t}(E_{\rm d}+E_{\rm s}) = 2\pi GM_{\rm
  p}\int_{R_{\rm in}}^{R_{\rm out}}(\Omega_{\rm
  s}-\Omega)\lambda\Sigma\mbox{d}R.
\end{equation}
This energy loss is positive definite, as it should be, because both
$\Omega_{\rm s}-\Omega$ and $\lambda$ change sign across the binary
orbit. The local dissipation term is thus
\begin{equation}
D_{\rm tid}(R) = \frac{GM_{\rm p}}{R}(\Omega_{\rm s}-\Omega)\lambda\Sigma.
\end{equation}
This term is only significant in a narrow range of radii at the disc
edges, which are hotter than a simple truncated disc. We include this
additional dissipation term in calculating the disc spectrum. 

This analysis assumes that the binary and disc orbits remain circular, and
so would predict a bright circular disc rim. In reality the binary is
likely to become somewhat eccentric. We discuss this point further in
the concluding section of this paper.

\subsection{Viscosity}

The final piece of physics needed in order to solve Eqs. (2-4) is a
prescription for the disc viscosity. We use the standard
$\alpha$-prescription \citep{shakura73} for a gas--pressure--dominated
disc:
\begin{equation}
\label{eq:viscosity}
\nu = \alpha c_{\rm s}^2/\Omega,
\end{equation}
where $c_{\rm s}=kT_{\rm c}/\mu m_H$ is the sound speed at the disc
midplane, $T_{\rm c}$ is the midplane temperature, $\mu = 0.67$ is the
mean molecular weight and $m_H$ is the mass of the proton. The
midplane temperature follows from solving the vertical radiative flux
equation for an optically thick disc, assuming the opacity $\kappa$ to
be either electron scattering or Kramers opacity, whichever the
larger, i.e.
\begin{equation}
\frac{9}{8}\nu\Sigma\Omega^2 = \sigma_{B}T_{\rm eff}^4
\end{equation}
where $\sigma_{\rm B}$ is Stefan-Boltzmann constant, and $T_{\rm eff}=T_{\rm
  c}/\tau^{1/4}$ is the effective temperature. The optical depth $\tau$ is
given by $\tau=\kappa \Sigma$ and the disc thickness $H$ is calculated from the
vertical hydrostatic balance including both gas and radiation pressure. This
setup, and in particular the requirement that the disc viscosity is
proportional to the gas pressure, Eq. (\ref{eq:viscosity}), ensures that the
disc is thermally stable. We take $\alpha=0.1$ throughout \citep{KingEtal07}.

\subsection{Gravitational instability and star formation}
\label{sec:gi}

The outer parts of AGN discs can be gravitationally unstable beyond $\sim$
0.01-0.1 pc \citep{kolikalov79,LB03b,goodman03,lodatoNC,KingPringle07}.  The
gravitational stability parameter $Q$ \citep{toomre64} is defined as
\begin{equation}
Q=\frac{c_{\rm s}\kappa_\Omega}{\pi G\Sigma},
\label{toomre}
\end{equation}
where $\kappa_\Omega$ is the epicyclic frequency. If $Q \lesssim 1$ the disc
is unstable. Further evolution of the disc is controlled by the ratio of
the cooling timescale, $t_{\rm cool}$, to the dynamical timescale
$\beta=t_{\rm cool}\Omega$. For $\beta\lesssim 1$ the disc fragments into
gravitationally bound objects \citep{gammie01,RLA05}. For longer cooling
times, $\beta\gtrsim 3$, the instability saturates at a finite amplitude and
the disc displays a long-lived spiral structure \citep{LR04,LR05}. In
protostellar accretion discs the parameter $\beta$ is generally much larger
than unity \citep{Rafikov05}. However in AGN discs the opposite occurs
(i.e. $\beta \ll 1$), and these discs are expected to fragment rather
easily \citep{goodman03,nayak06,nayak07,Levin07}.

The evolution of a fragmenting $\beta \ll 1$ disc is uncertain. The
prescription of a constant cooling time throughout the collapse of a
fragmenting region is an oversimplification. If the collapse of
self--gravitating fragments is not prevented, most of the mass quickly
(almost dynamically) ends up in stars \citep[e.g.,][]{nayak07} and
this might inhibit accretion altogether. Conversely, the energy input
arising from the star formation process (either via protostellar
accretion or from nuclear burning in the newly formed stars) could
keep the disc close to marginal stability
\citep{goodman03,ThompsonEtal05,nayak06}.

In this paper we explore this latter possibility. In particular, at each
timestep we compute the gravitational stability parameter $Q$ (equation
\ref{toomre}) for every radial bin in the disc. If $Q < 1$, star formation
is allowed. We define the local star formation rate per unit area, $
\dot{\Sigma}_{\rm sf}$. The appropriate value of this rate is found by
requiring that the corresponding energy liberation rate is high enough to
keep the disc in the state of marginal stability, assumed to be $Q = 1$. The
internal heating of the disc is then dominated by the feedback from star
formation, and hence the energy equation becomes
\begin{equation}
\label{eq:SF}
\epsilon \dot{\Sigma}_{\rm sf} c^2 = \sigma_{B}T_{\rm eff}^4\;.
\end{equation} 
The free parameter of the model, $\epsilon$, is the efficiency of
mass--to--energy conversion in star formation. The star formation rate
per unit surface is added as a sink term on the rhs of
Eq. (\ref{eq:diffplanet}).

A reasonable estimate of $\epsilon$ is obtained by calculating the
total amount of energy liberated by nuclear reactions and supernovae
assuming a Salpeter IMF from $0.1\msun$ to $100\msun$. This yields
$\epsilon \sim10^{-3}$ \citep{ThompsonEtal05}. However $\epsilon$
could in principle be lower or higher than this value. A higher value
$\epsilon \sim 0.01$ could be reached if the IMF of stars is
top--heavy, as found for the star formation event in the central $0.5$
pc of the Milky Way \citep{NS05,PaumardEtal06}. A lower value of
$\epsilon$ might be appropriate if the stellar disc thickens quickly and
adopts a geometrically thicker configuration than the gaseous disc
\citep{nayak07}. Most of feedback energy would then be escaping from
the disc rather than being deposited in the midplane.

Another important effect is the increased pressure in the disc
\citep{ThompsonEtal05} from winds and supernovae. However we limit our
attention here to the energy effects of star formation only, and leave
pressure effects to a future paper.

\section{Time-dependent calculations}

Here we describe the results of our simulations. We have run two sets, first
without star formation (Section~\ref{sec:noSF}), and then taking feedback from
star formation into account (Section~\ref{sec:withSF}).

\subsection{Initial conditions and disc setup}

We solve the set of equations (2)-(4) using standard techniques (see,
for example, \citealt{LC04}). We use a logarithmically spaced radial
grid extending from the innermost stable orbit at $6R_{\rm g}$, where
$R_{\rm g}=GM_{\rm p}/c^3$ is the primary gravitational radius, to an
outer radius of $20a_0$, where $a_0$ is the initial binary
separation. We typically use 100 mesh points. We use a zero torque
boundary condition at both our innermost grid point (i.e. close to the
primary black hole) and at our outermost one, at 20$a_0$. The
outer boundary condition does not really play a role since the disc
never extends that far out.

At each time during the simulation we compute the effective temperature
$T_{\rm eff}(R)$ at the disc surface, including both the standard viscous
dissipation term and the additional tidally induced term described
above. From $T_{\rm eff}$ we calculate the disc spectral energy distribution
and the total luminosity by integrating a series of blackbody spectra with
the appropriate radial dependent temperature over the disc surface.

All our calculations start with similar initial conditions. We take
the primary black hole to have a mass of $M_{\rm p}=10^8M_{\odot}$.
We have run a number of simulations with different secondary mass,
total disc mass $M_{\rm d}$, and different initial binary separation
$a_0$. The disc is initialised with a constant surface density over a
narrow range of radii between $0.8a_0$ and $2a_0$, where the density
normalisation is chosen to give the required total disc mass. This is
intended to give a simple realisation of the concept that both black
hole and gas arrive at small radii as part of the same merger
process. Although clearly an idealisation, we regard such initial
configuration more plausible than the standard assumption that the
black hole is somehow parachuted into an accretion disc that has already
had time to relax into a steadily accreting state. We have considered
the two initial separations $a_0= 0.01$~pc and $a_0=0.05$~pc. During a
gas--rich merger of two galaxies each hosting a black hole, the two
holes can typically reach a distance of $\lesssim 0.1$~pc by dynamical
friction against the stellar and gaseous background
\citep{escala05,dotti06}. Thus our initial conditions are rather
optimistic as to how far in a black hole can progress. Even so we find
problems with merging on reasonable timescales even from such small
initial separations. This must imply stringent limits on the
possibility of gas--driven coalescences.

\subsection{Discs without star formation}
\label{sec:noSF}

\begin{figure*}
  \begin{center}
      \includegraphics[width=0.3\textwidth]{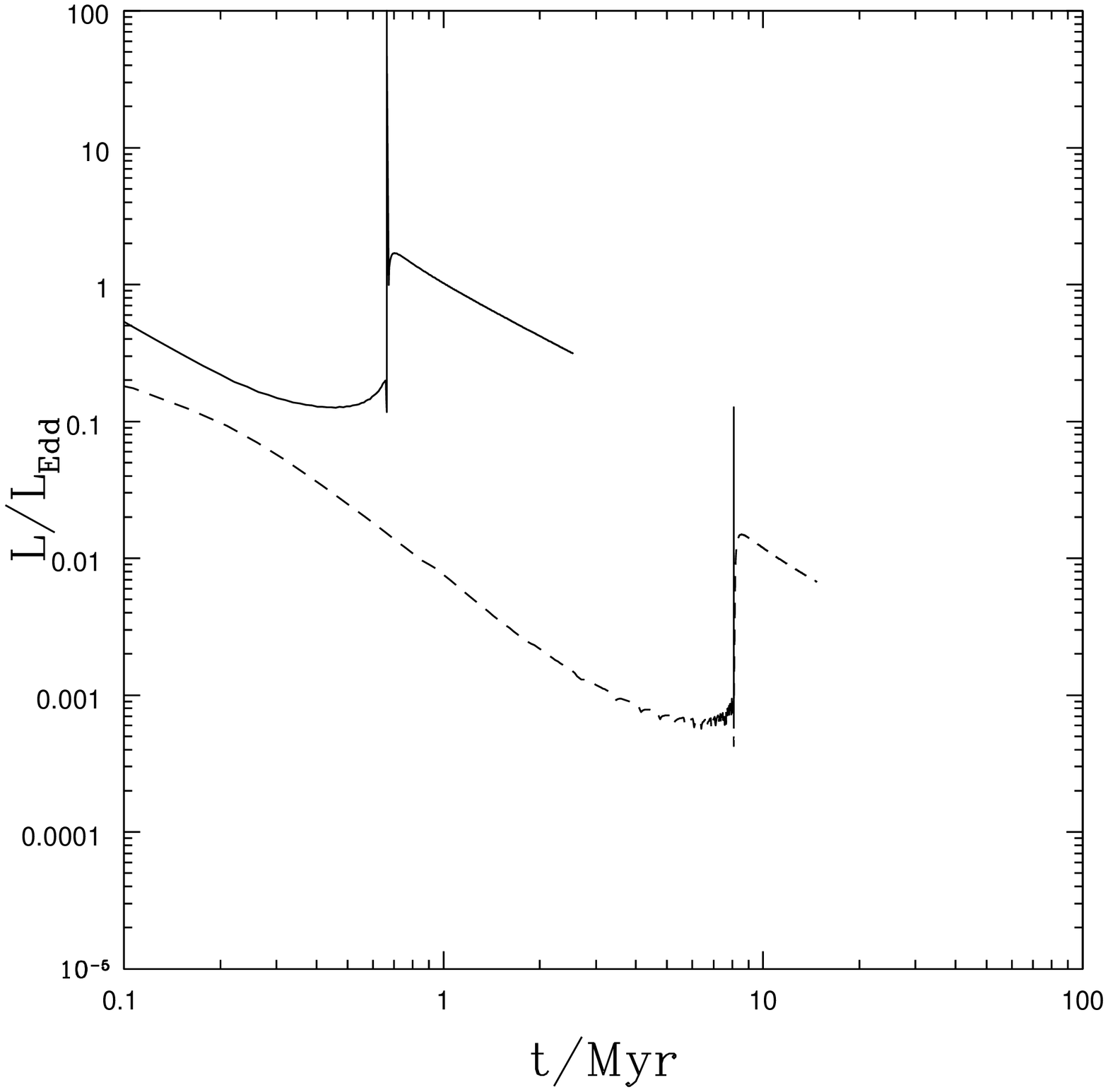}
      \includegraphics[width=0.3\textwidth]{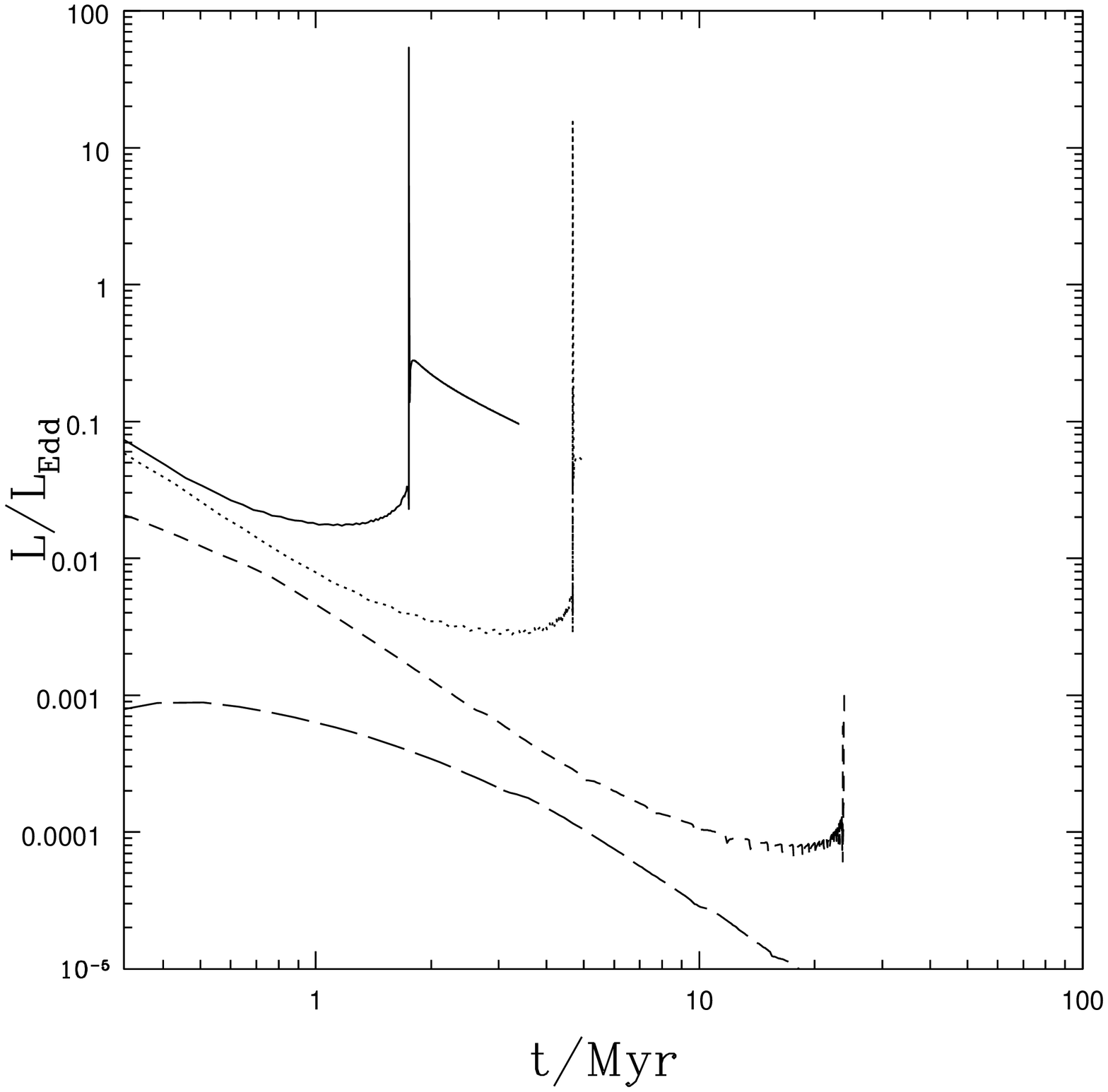}
      \includegraphics[width=0.3\textwidth]{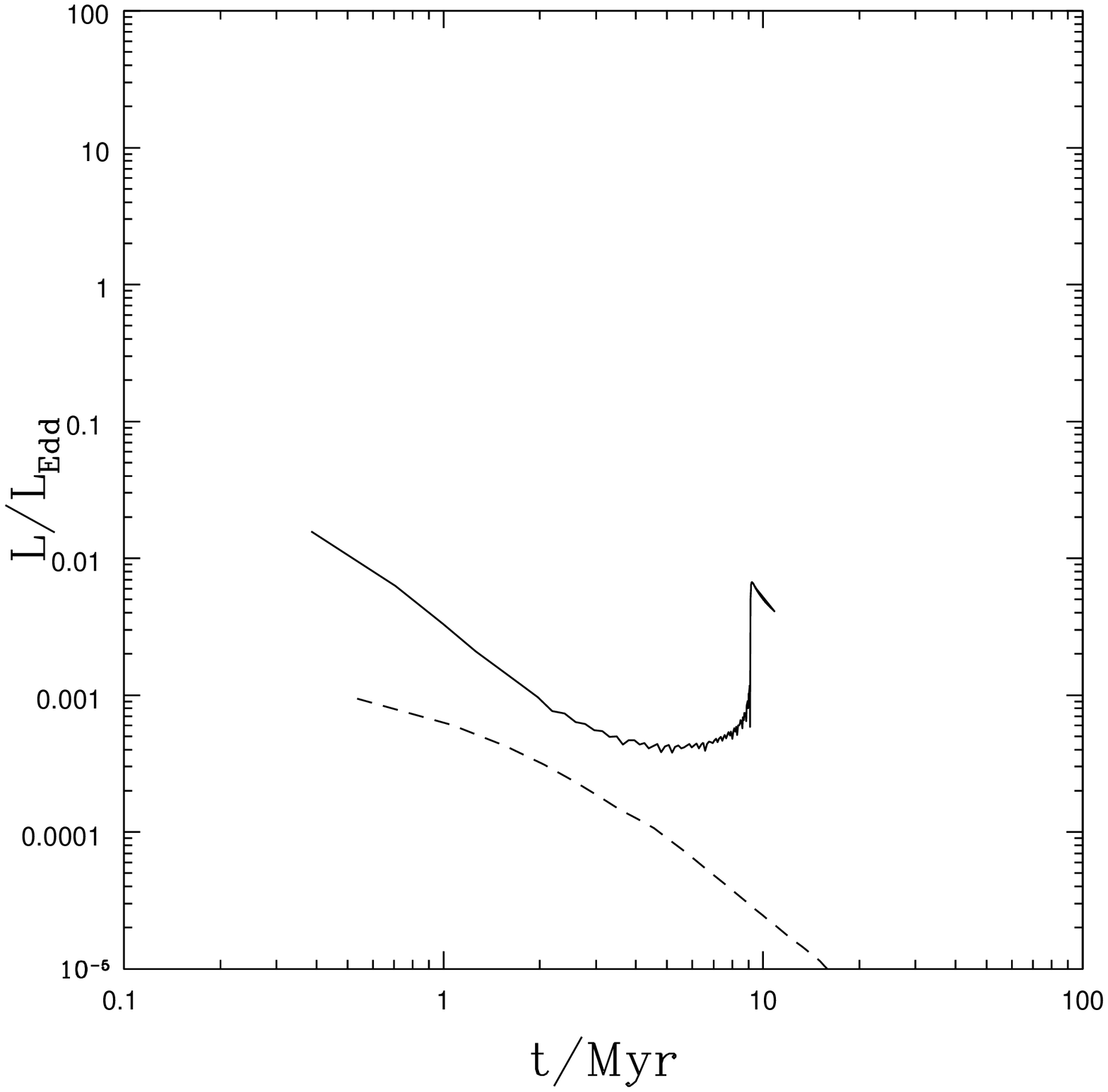}
      \caption{Luminosity evolution of circumbinary discs for an
        initial separation of $a_0 = 0.01$ pc.  The secondary/primary
        mass ratio is $M_{\rm s}/M_{\rm p}= q =$ 0.3 (left panels), 0.1
        (middle panel) and 0.01 (right panel). For each panel, the
        various lines refer to different disc masses: $M_{\rm
          d}/M_{\rm s}= 1$ (solid lines), 0.5 (dotted line), 0.1
        (short-dashed line) and 0.01 (long-dashed line).}
	\label{fig:luminosity}
  \end{center}
\end{figure*}

\begin{figure*}
  \begin{center}
      \includegraphics[width=0.3\textwidth]{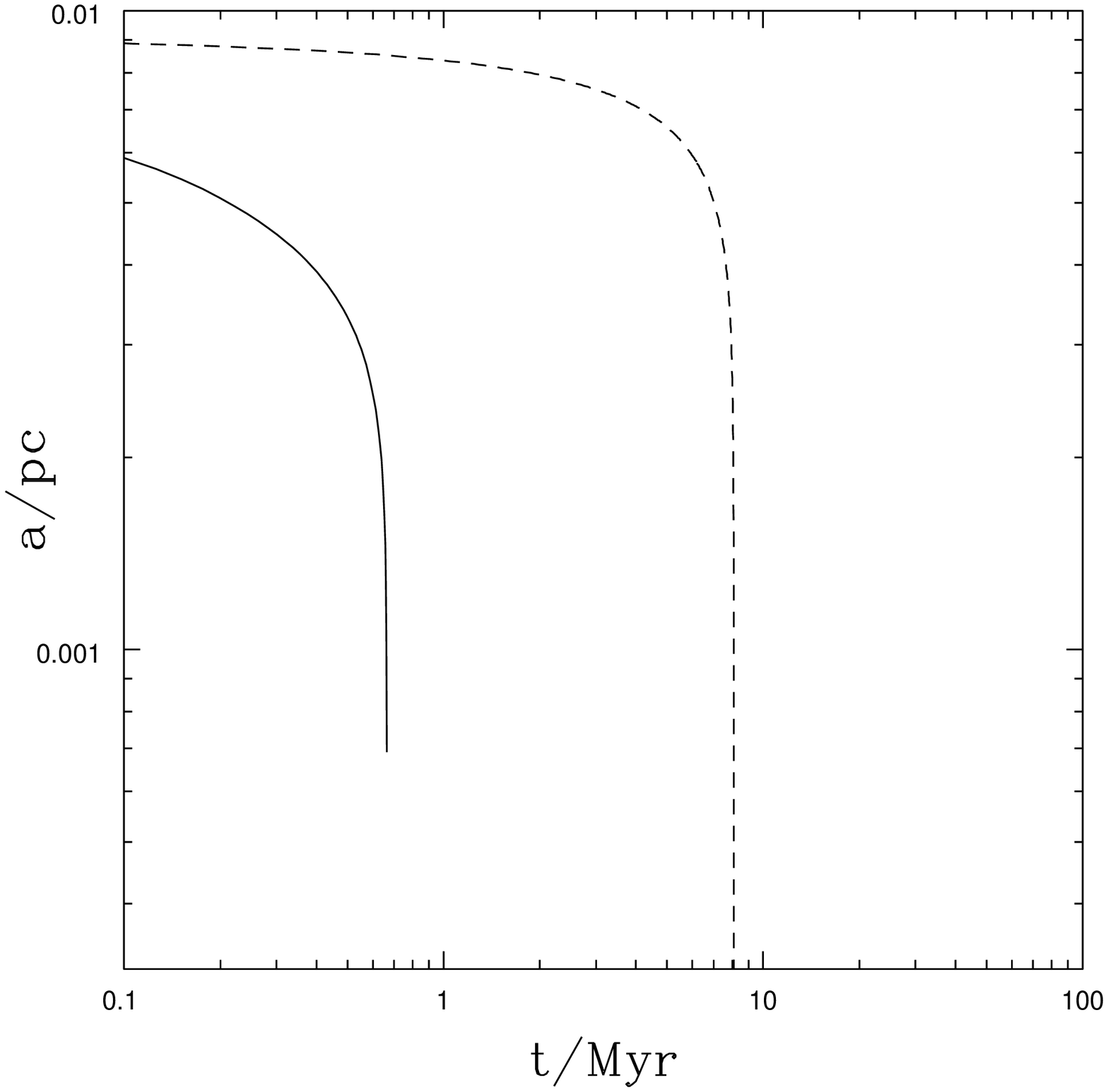}
      \includegraphics[width=0.3\textwidth]{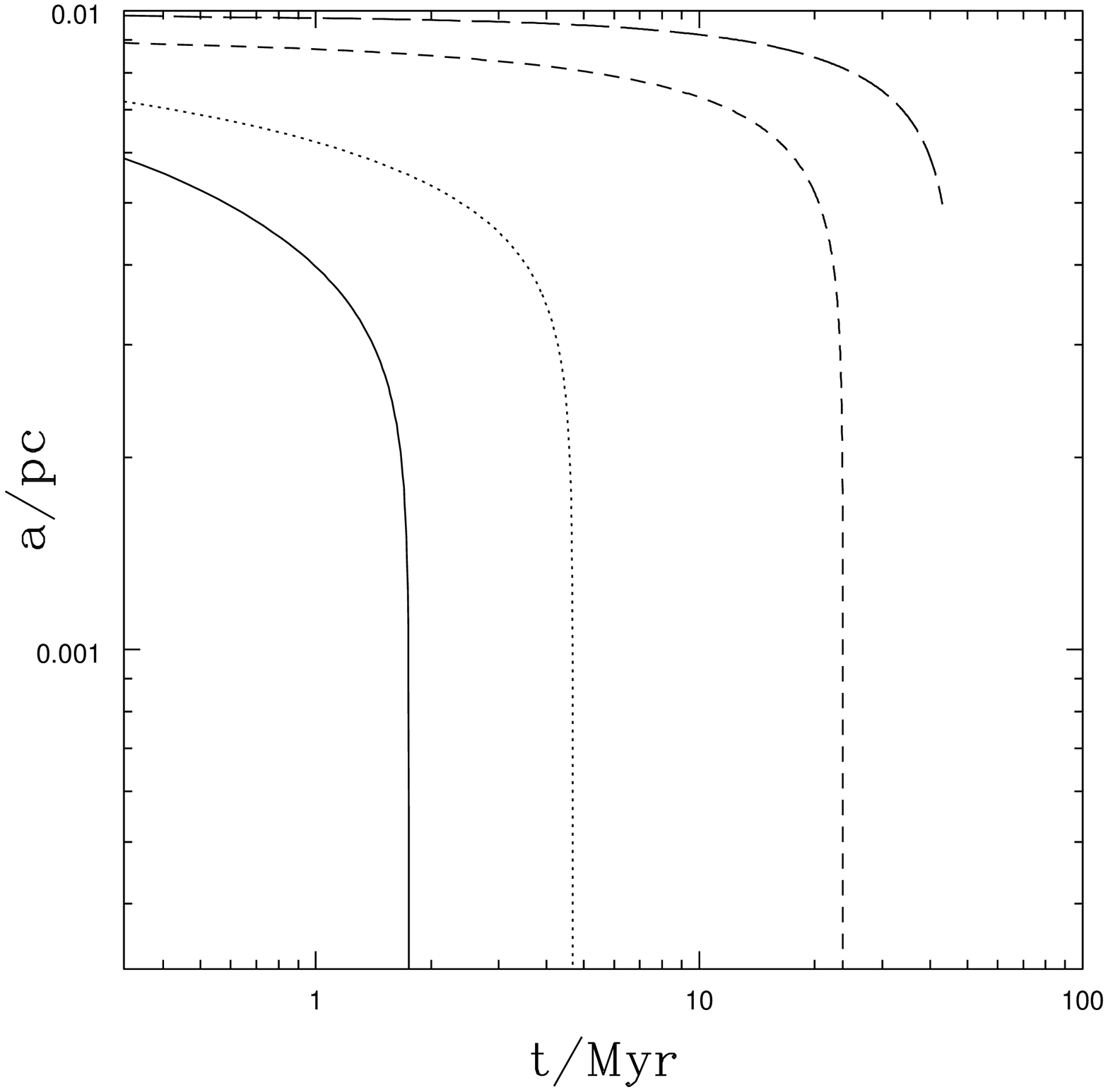}
      \includegraphics[width=0.3\textwidth]{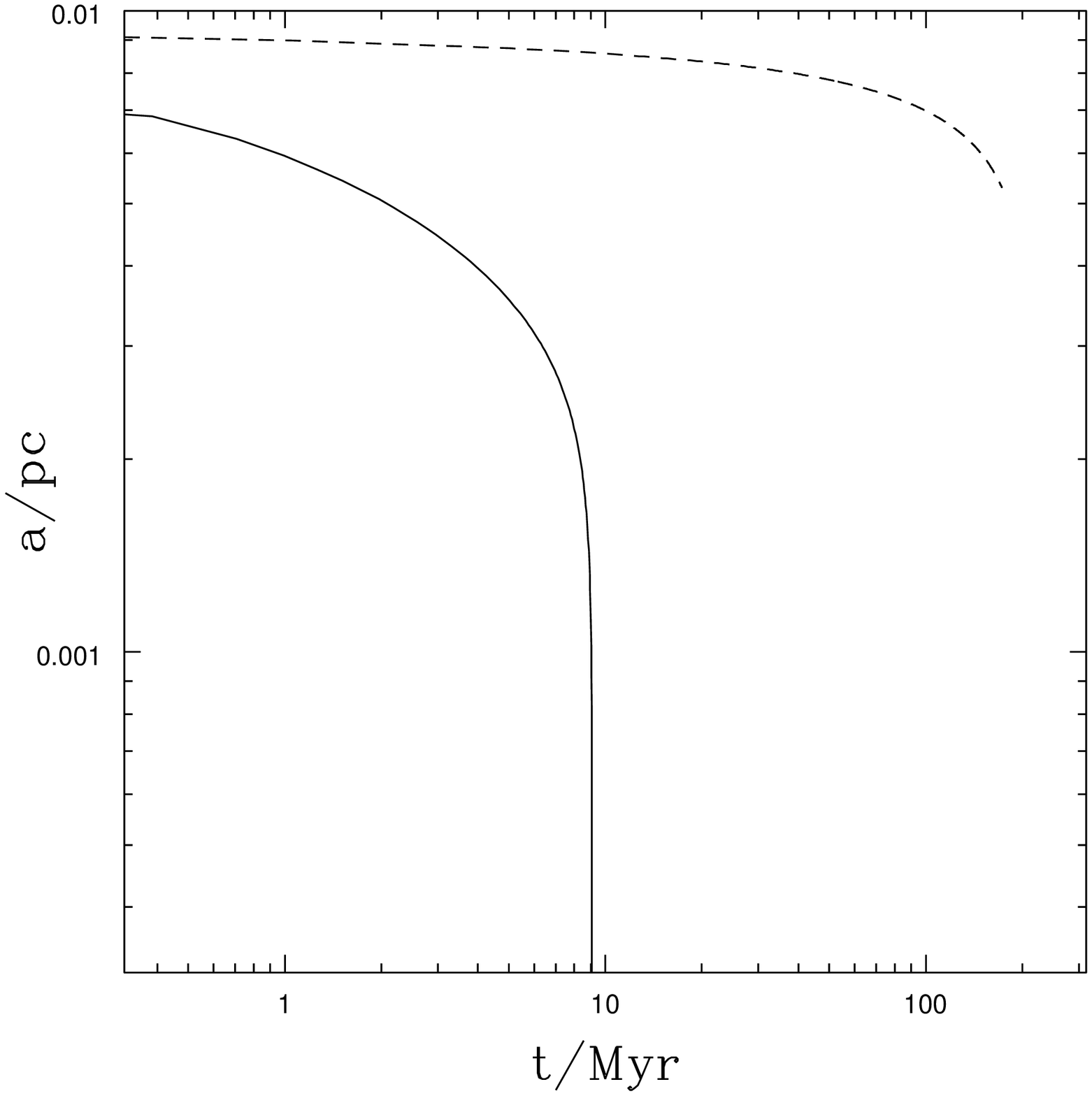}
      \caption{Evolution of the binary separation $a$ for the initial
        value of $a_0 = 0.01$ pc. The three plots have the same
        parameters as the corresponding plots in
        Fig. \ref{fig:luminosity}. Similarly, the line styles are the
        same as in Fig. \ref{fig:luminosity}.}
	\label{fig:separation}
  \end{center}
\end{figure*}

In the first set of simulations we neglect the effects of star formation and 
allow the disc to become unstable ($Q<1$) without taking any further
action. Although obviously unrealistic, these simulations are useful for
comparing with those where star formation is included, and also to illustrate
the state of the disc at decoupling, i.e. at the point where the binary
coalescence starts to be driven by gravitational wave emission rather than
by disc torques, as a function of the main disc parameters. By insisting
that no gas is removed by star formation we overestimate the degree to which
the gas can facilitate a black hole merger.

Our results should be compared with those of \citet{milosavljevic05},
who consider the same general setup but do not follow the
time-dependent evolution of the system. Instead they simply truncate
the disc at the outer radius where it becomes gravitationally
unstable. We therefore briefly summarise their main findings. They
argue that, contrary to what happens when the mass ratio ratio $q\ll
1$ \citep{natarajan02}, inner disc evolution for non--extreme mass
ratios is much faster than the binary orbital evolution, so that after
a short time the only gas present in the system is a circumbinary
disc. We are interested in the state of the disc at decoupling, i.e.
at the moment where binary evolution driven by gravitational wave
losses becomes faster than the disc viscous evolution and the disc
inner edge cannot follow the secondary black hole as it spirals in.
\citet{milosavljevic05} assume the circumbinary disc at decoupling to
be well described by a constant--$\dot{M}$ steady state solution,
truncated at the inner edge. Using the requirement that the disc is
thin at the inner edge, they argue that $\dot{M}<\dot{M}_{\rm Edd}$,
where $\dot{M}_{\rm Edd}=L_{\rm Edd}/\eta c^2$ is the Eddington
accretion rate \emph{evaluated at the innermost stable orbit}, $L_{\rm
  Edd}$ is the Eddington luminosity and $\eta$ is the accretion
efficiency. The resulting spectral energy distribution is therefore
approximately given by a standard multi-colour blackbody spectrum, but
lacking the high--energy component and therefore peaking at
optical--IR wavelengths. After decoupling the binary torques are
removed and the disc is free to flow to the bottom of the potential
well, producing in a bright flare, and thus recovering the
high--energy part of the standard disc spectrum. As we shall see, this
is not the complete picture.

\subsubsection{Initial binary separation equal to 0.01 pc}

We consider first the case where the initial binary separation is $a_0 =
0.01$ pc, although the general behaviour of the system is similar for all
the simulations.  At this distance from a $10^8M_{\odot}$ primary, the
timescale for a merger induced purely by gravitational radiation is $t_{\rm
gw}\approx 2\times 10^7q^{-1}$ yrs.

The luminosity as a function of time for all our simulations is shown in
Fig. \ref{fig:luminosity}, where the left panel refers to $q= 0.3$,
the middle panel refers to $q=0.1$ and the right panel refers to
$q=0.01$. The various lines indicate different values for the total disc
mass $M_{\rm d}$ (see caption for details). We consider a range of cases
varying from a disc mass comparable to the secondary mass $M_{\rm d}/M_{\rm
s} = 1$ to one where it is much smaller $M_{\rm d}/M_{\rm s} = 0.01$. When
the disc mass is much larger than the secondary mass, the secondary black
hole is simply swept in to the central black hole by the accretion flow. 

We start with roughly half of the gas inside the orbit of the
secondary black hole and half outside. Initially, the matter inside
the binary orbit accretes on to the primary black hole, resulting in a
relatively bright AGN, where the luminosity obviously scales with the
total disc mass. For the choice of parameters made here the peak
luminosity is at most of order the Eddington luminosity for our most
massive disc. The inner disc evolves faster then the binary orbit and
the surface density and the disc luminosity drops rapidly. However, a
low density disc is still present at decoupling in all of our
simulations. Clearly, the amount of mass left over in the inner disc
at decoupling is strongly model--dependent. However, the presence of a
small, but non--negligible, mass in the inner disc might have important
observational consequences. As explained above, even if such a low density disc does not provide a significant contribution to the SED, it can be squeezed when angular momentum loss by gravitational radiation becomes dominant, producing a sudden burst of luminosity. After this, the outer disc can finally flow to the innermost region and provide a longer--lasting AGN--like appearance, even though the post--merger luminosity is generally sub--Eddington.

\begin{figure}
  \begin{center}
      \includegraphics[width=0.45\textwidth]{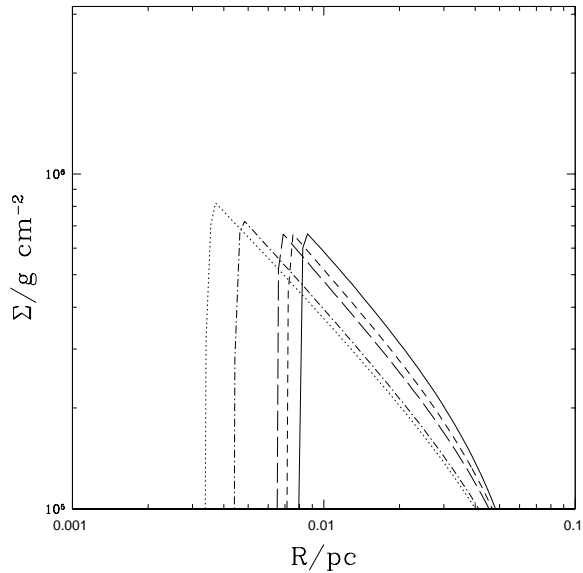}
      \caption{Surface density of the outer disc for the case $q=1$,
        $M_{\rm d}/M_{\rm s} = 1$. The lines refer to $t = 0.5$ Myr
        (solid line), 0.75 Myr (short-dashed line), 1Myr (long-dashed
        line), 1.5 Myr (dot-dashed line) and 1.6 Myr (dotted
        line). The last two cases refer to epochs when the binary orbit
        accelerates because of gravitational wave emission
        (cf. Fig. \ref{fig:separation}), and the torque on the inner
        disc is removed.}
	\label{fig:sigmavstime}
  \end{center}
\end{figure}

Fig.~\ref{fig:separation} shows the evolution of the binary
separation with time. The initial evolution is rather slow, being driven
mostly by the disc torques. It suddenly accelerates when the gravitational
radiation induced evolution takes over and the outer disc decouples from the
binary evolution. We see from Fig.~\ref{fig:separation} that the radius at
which the disc decouples increases with decreasing disc mass. A similar
result was also found by \citet{milosavljevic05}, who parameterised the disc
mass terms of the mass flux at the disc inner edge (see below). Note that
for disc masses smaller than roughly 0.1 of the secondary mass, the merger
time becomes comparable with the gravitational radiation timescale,
indicating that the disc gives only a limited contribution towards
accelerating the merger.

Figure \ref{fig:sigmavstime} shows a series of snapshots of the
surface density in the outer disc for the case where $q=0.1$ and
$M_{\rm d}/M_{\rm s} = 1$ (which corresponds to the solid line in the
middle panels of Figs. \ref{fig:luminosity} and
\ref{fig:separation}). The lines refer to $t = 0.5$ Myr (solid line),
0.75 Myr (short--dashed line), 1Myr (long--dashed line), 1.5 Myr
(dot--dashed line) and 1.6 Myr (dotted line). This clearly shows that
in the situation considered in this paper the surface density at the
inner edge does not grow with time (unlike the \citealt{ivanov99}
solution). The eventual growth of $\Sigma$ at the inner edge only
occurs at late times, when the binary orbit accelerates due to
gravitational wave emission (cf. Fig. \ref{fig:separation}) and the
tidal torque is effectively removed.

\begin{figure*}
  \begin{center}
      \includegraphics[width=0.45\textwidth]{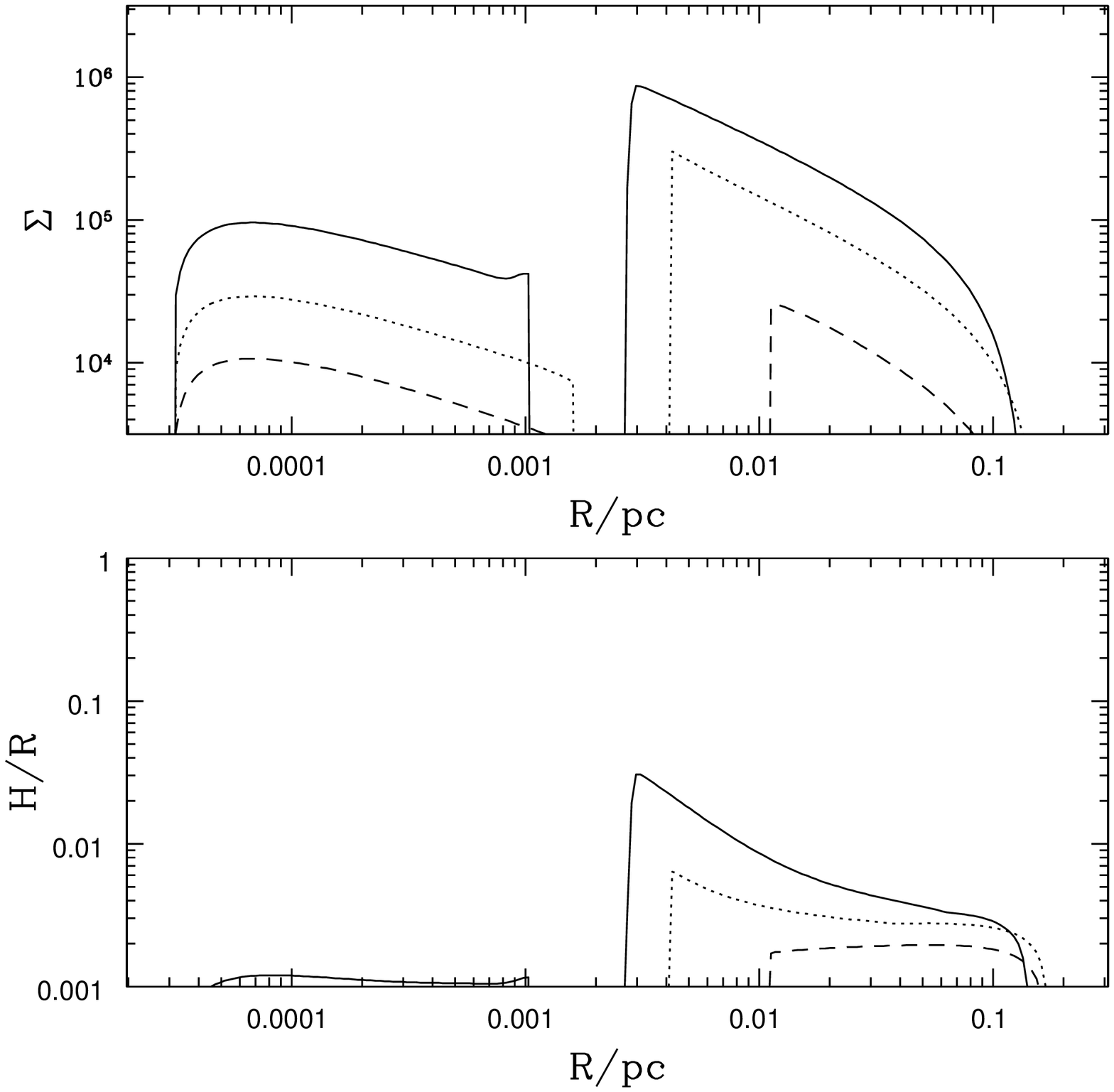}
      \includegraphics[width=0.45\textwidth]{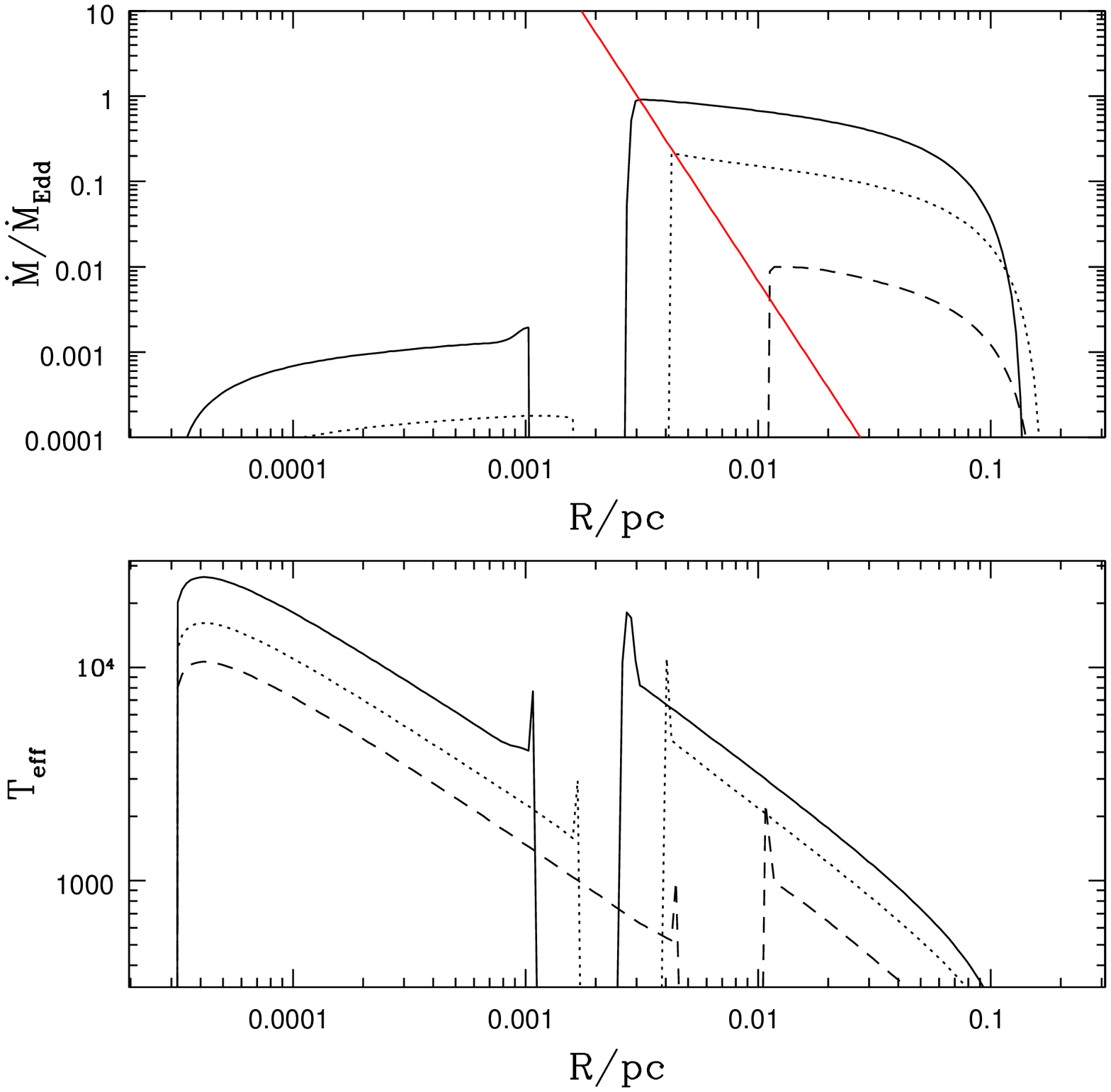}
      \caption{Properties of the disc at decoupling, i.e. when the
        binary evolution via gravitational wave radiation becomes
        faster than viscous evolution at the secondary's orbit. The
        initial separation was $a_0 = 0.01$ pc. The secondary/primary
        mass ratio is $q=0.1$. The different lines refer to different
        values of the initial disc mass, with the same notation as
        Fig. \ref{fig:luminosity}. Shown here is: the disc surface
        density profile in CGS units (upper left), the mass flux
        through the disc in units of the Eddington rate at the
        innermost stable orbit (upper right), the aspect ratio $H/R$
        (lower left) and the effective temperature in Kelvin (lower
        right).}
	\label{fig:decoupling}
  \end{center}
\end{figure*}

Figure~\ref{fig:decoupling} shows a number of properties of the disc
at decoupling, which is the moment at which the binary separation
evolution becomes faster than the disc viscous time close to the
binary orbit. The top panels show the disc surface density profile
(left) and the profile of the local mass flux in the disc $\dot{M} =
2\pi\nu\Sigma$ in units of the Eddington rate at the innermost stable
orbit $\dot{M}_{\rm Edd}$, where we assume mass to energy conversion
efficiency of $\eta=0.1$ (we use this scaling for ease of comparison
with the results of \citealt{milosavljevic05}). The bottom plots show
the aspect ratio $H/R$ (left) and the effective temperature profile
(right). These plots refer to $q=0.1$ and the three lines
correspond to to $M_{\rm d}/M_{\rm s} = $ 1 (solid line), 0.5 (dotted
line) and 0.1 (dashed line).

There are several interesting features in these plots. We first
discuss the radial profiles of $\dot{M}$. The main thing to note is
that the outer circumbinary disc does {\it not} have a constant
$\dot{M}$, as a standard steady accretion disc model would
require. The tidal torque at the inner edge of the circumbinary disc
changes its overall structure, such that if a steady state is reached
the mass flux scales as $\dot{M}\propto R^{-1/2}$. This kind of
profile is typical of viscous disc solutions in the presence of a
torque at the inner edge \citep{pringle91,clarkesyer95,ivanov99}, and
as we will discuss below it affects the shape of the spectral energy
distribution of the disc.  We stress that the choice of scaling
$\dot{M}$ to the inner Eddington rate is motivated by convenience, and
the mass flux can significantly exceed the reference value (the
Eddington value for the innermost stable orbit) without the luminosity
becoming super--Eddington, since the disc is truncated at a radius
much larger than the innermost stable orbit. \citet{milosavljevic05}
still argue that the limit $\dot{M}/\dot{M}_{\rm Edd}<1$ is required
in order to keep the disc thin at its inner edge, at least for
relatively large $q$. Although our most massive disc does
have $\dot{M}/\dot{M}_{\rm Edd}\approx 1$ at its inner edge, we find
that the disc aspect ratio $H/R$ is $\ll 1$ (see bottom left panel),
therefore in principle allowing a much larger inner mass flux without
violating the thin disc requirement. The straight (red) line in the
top right panel of Fig. \ref{fig:decoupling} shows the scaling between
the inner edge of the circumbinary disc at decoupling and the mass
flux predicted by \citet{milosavljevic05}, where the normalisation has
been chosen to match our results. Although our simulations do
reproduce this scaling quite well, our results imply a binary
separation at decoupling larger by a factor two than
\citet{milosavljevic05}. At decoupling we have $a= 335,~520,~1400
R_{\rm g}$, for the three cases shown in Fig. \ref{fig:decoupling},
i.e. $M_{\rm d}/M_{\rm s} = $1, 0.5 and 0.1, respectively.

The top left panel in Fig. \ref{fig:decoupling} shows the surface density
profile at decoupling. Here we note that although the inner circumprimary
disc has been significantly depleted in mass and its surface density is thus
low, it has still not disappeared. As emphasised above, even a small amount of
mass in the circumprimary disc produces a bright burst as it is squeezed by
the final evolution of the binary orbit, producing the sudden peak in the
light curves plotted in Fig. \ref{fig:luminosity}. The bottom right panel in
Fig. \ref{fig:decoupling} shows the effective temperature profile at
decoupling. Here the most significant feature is the presence of two bright
spots at the edges of the gap around the secondary. This results from the
extra dissipation term discussed in Section 3.3, and provides a significant,
and previously neglected, contribution to the overall spectral energy
distribution.

\begin{figure}
  \begin{center}
      \includegraphics[width=0.45\textwidth]{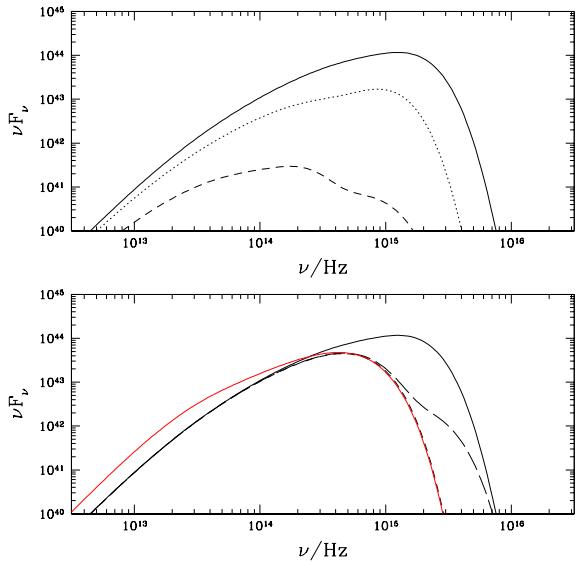}
      \caption{Top: SED of the disc (in CGS units) at decoupling for
        various disc masses: $M_{\rm d}/M_{\rm s}$ =1 (solid line),
        0.5 (dotted line) and 0.1 (dashed line). The initial
        separation was $a_0 = 0.01$ pc. Bottom: Contributions to the
        overall SED for disc mass $M_{\rm d}/M_{\rm s}=1$ shown in the
        top plot. The solid line indicates the full spectrum; the
        short--dashed line shows the contribution of only the outer
        disc, without including the extra tidal dissipation term; and
        the long--dashed line includes the emission of the inner disc
        but still does not include the hot gap edges. Comparing the
        long--dashed and solid lines thus shows the importance of the
        hot--gap emission at high radiation frequencies.  For
        reference, the red line shows the SED of a standard
        constant--$\dot{M}$ accretion disc: its spectrum is shallower
        than the one found here, which more closely approximates the
        structure of a {\it decretion} disc.}
	\label{fig:sed}
  \end{center}
\end{figure}

The resulting spectral energy distribution (SED) for the three cases
of Fig. \ref{fig:decoupling} are shown in the top panel of
Fig. \ref{fig:sed}. These are calculated as multicolour blackbody
spectra for the appropriate effective temperature profiles. The
general shape is similar to that predicted by \citet{milosavljevic05},
although there are significant differences. The disc is very bright at
optical/IR wavelengths and lacks the high energy emission from the
inner disc. The inner disc emission is generally negligible because of
the low density of this disc, although some high energy emission from
the inner disc can be seen for the lowest disc mass (dashed
line). The bottom plot in Fig. \ref{fig:sed} shows the contribution to
the total SED of the various disc components, for $M_{\rm
  d}/M_{\rm s}=1$. The short--dashed line shows the contribution of
the outer circumbinary disc only, without including the extra
dissipation term at the gap edges. The long--dashed line shows the
spectrum arising from the whole disc, but neglecting again the
dissipation at the gap edges. This emphasises the contribution of the
inner disc. The solid line shows the total SED, including the emission
from the gap edges, which in turn result in a significant contribution
at relatively higher energies. The red line, for comparison, shows the expected SED of a
truncated disc with constant $\dot{M}$. This is the shape of the SED
as predicted by \citet{milosavljevic05}. As one can see, the
long--wavelength part of the SED of a constant $\dot{M}$ disc is
shallower than the one we calculate. The steepening of the SED in the
presence of a massive satellite had been discussed by
\citet{clarkesyer95}. Then we have $\nu F_{\nu} \propto \nu^{12/7}$,
rather than the value of $\nu F_{\nu} \propto \nu^{4/3}$ found for a
steady accretion disc.

Note that the frequency range of the hot edge emission is comparable
to that of the inner disc. Indeed Fig. \ref{fig:decoupling} shows that
the effective temperature of the edge is similar to that of the inner
disc. However, the area covered by the hot edges is much larger than
the inner disc and provides a much larger luminosity. The total
luminosity of the hot edges is roughly equal to half of the bolometric
luminosity. 

After the black holes coalesce and the tidal
torque at the inner edge is removed, the gas will eventually freely
flow to the inner disc, providing a luminous high energy component to
the SED, which would gradually flatten to reach a standard
constant--$\dot{M}$ solution in the inner disc.

\subsubsection{Initial binary separation equal to 0.05 pc}

\begin{figure}
  \begin{center}
      \includegraphics[width=0.45\textwidth]{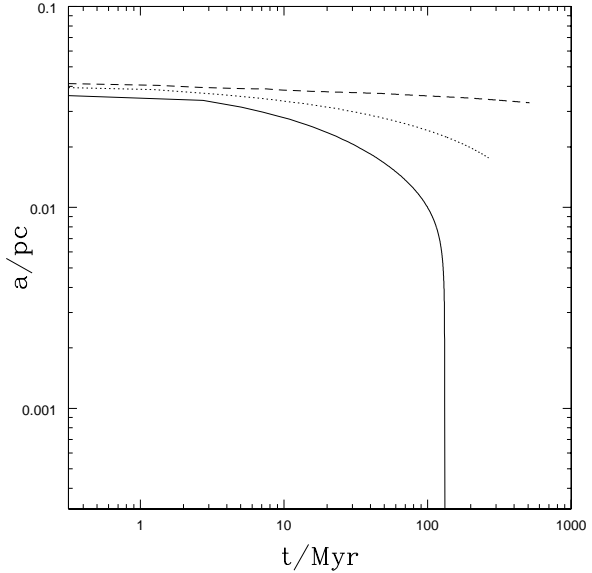}
      \caption{Evolution of the binary separation with time when the initial
        separation is taken to have the larger value of $a_0 = 0.05$ pc. The
        primary mass here is $10^8M_{\odot}$, the secondary/primary mass ratio
        is $q=0.1$ and the disc/secondary mass ratio is $M_{\rm d}/M_{\rm
          s}=1$ (solid line), 0.5 (dotted line) and 0.1 (dashed
        line). Only the more massive discs produce a merger within 1 Gyr.}
	\label{fig:a_pc5}
  \end{center}
\end{figure}

For a larger initial binary separation of $a_0 = 0.05$ pc, the
gravitational radiation timescale is far longer and would not produce
coalescence in a Hubble time without any further binary hardening
mechanism. Here again we take the primary black hole mass to be
$10^8M_{\odot}$. The mass ratio between the secondary and the primary
is $q=0.1$ and we consider, as in the previous section, different
values for the disc/secondary mass ratio. Fig. \ref{fig:a_pc5} shows
the evolution of the binary separation for $M_{\rm
  d}/M_{\rm s} = 1$ (solid line), 0.5 (dotted line) and 0.1
(dashed line). Here again, we see the same trend as observed in
for the smaller initial separation $a_0 = 0.01$ pc. We note that
disc torques are able to induce coalescence within a Hubble time only
when the disc mass is comparable to the secondary black hole. At the
end of our simulations ($\sim 1$~Gyr) the binary had merged only for
$M_{\rm d} = M_{\rm s}$. The final state of the disc at decoupling
follows the same trend as before.

\subsection{The effects of star formation}
\label{sec:withSF}

From the analysis of the results of the previous section, we may draw
the following general conclusion. Even starting from an initial
separation of $a_0 = 0.05$ pc, a circumbinary disc is able to bring
the SMBH binary to coalescence in a reasonable time only if the disc
mass is of the order of, or slightly smaller than, the secondary black
hole mass. However, in the previous section we did not include the
effects of gravitational instability and simply allowed the disc to
have a value of $Q$ smaller than one in its outer parts, beyond
$\approx 0.05$ pc. Here, we include the effects of gravitational
instability by requiring that, whenever the disc is unstable, it is
kept in a state of marginal stability by the feedback induced by star
formation.  We thus remove from the calculation the amount of mass
needed to be converted into stars to provide such feedback, as
detailed in Section \ref{sec:gi}. The effect of removing gas from the
disc is to reduce the efficacy of the disc in merging the black holes.

\begin{figure}
  \begin{center}
      \includegraphics[width=0.45\textwidth]{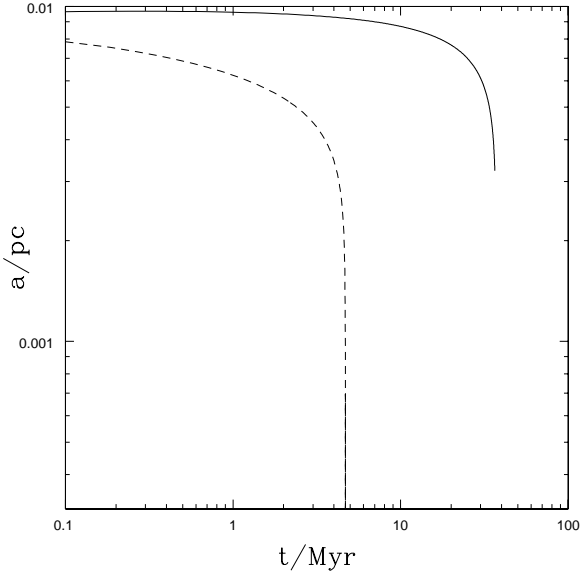}
      \caption{Evolution of the binary separation with time in
        calculations with (solid curve, see \S \ref{sec:withSF}) and
        without (dashed, see \S \ref{sec:noSF}) star formation in the
        disc. The latter case is the one shown in
        Figures~\ref{fig:separation} and \ref{fig:decoupling}. The
        primary mass is $10^8M_{\odot}$, the secondary/primary mass
        ratio is $q=0.1$, the disc/secondary mass ratio is $M_{\rm
          d}/M_{\rm s}=0.5$ and the initial separation is $a_0
        =0.01$~pc. Note that removal of gas by star formation
        significantly slows down binary merging.}
	\label{fig:a_SF}
  \end{center}
\end{figure}

We first consider the case where the secondary/primary black hole mass
ratio is $q=0.1$, the initial mass ratio between the disc and the
secondary is $M_{\rm d}/M_{\rm s}=0.5$ and the initial binary
separation is 0.01 pc.  We also set the star formation feedback
efficiency to $\epsilon = 10^{-3}$. In Fig. \ref{fig:a_SF} we show the
resulting evolution of the binary separation (solid line) and that in
the corresponding calculation without allowing for star formation
(dashed line).  Star formation significantly reduces the ability of
the disc to induce the merger. This is because star formation depletes
the gas reservoir in the outer parts of the disc, reducing its surface
density, and so significantly reduces the disc density just outside
the secondary orbit.

\begin{figure}
  \begin{center}
      \includegraphics[width=0.45\textwidth]{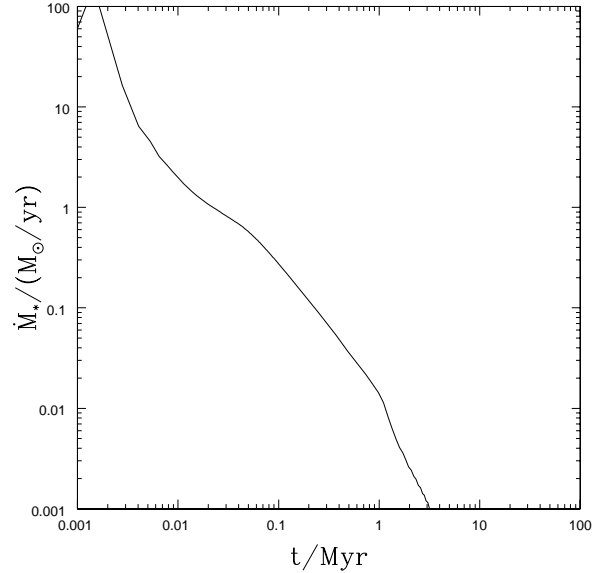}
      \caption{Total star formation rate for the simulation shown also in
        Fig. \ref{fig:a_SF}. Over the entire course of the simulation the
        total amount of gas turned into stars is $\approx 2~10^6M_{\odot}$.}
	\label{fig:SF}
  \end{center}
\end{figure}

\begin{figure}
  \begin{center}
      \includegraphics[width=0.45\textwidth]{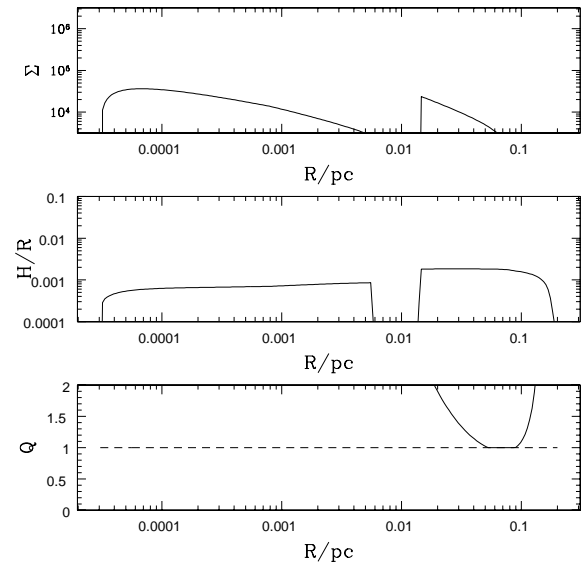}
      \caption{Disc properties after 6 Myr. At this time star
        formation has considerably depleted the disc mass and little
        disc evolution has taken place. Because of this the results
        are relatively independent of initial disc mass. The initial
        separation was $a_0 = 0.01$ pc, the secondary/primary black
        hole mass ratio is $q=0.1$, the disc mass is $M_{\rm d}/M_{\rm
          s} = 0.5$ and the star formation feedback efficiency is
        $\epsilon = 10^{-3}$. Top panel: disc surface density in cgs
        units. Middle panel: aspect ratio $H/R$. We see that typically
        $H/R \approx 10^{-3}$. Bottom panel: gravitational stability
        parameter $Q$.}
	\label{fig:combine_SF}
  \end{center}
\end{figure}

\begin{figure}
  \begin{center}
      \includegraphics[width=0.45\textwidth]{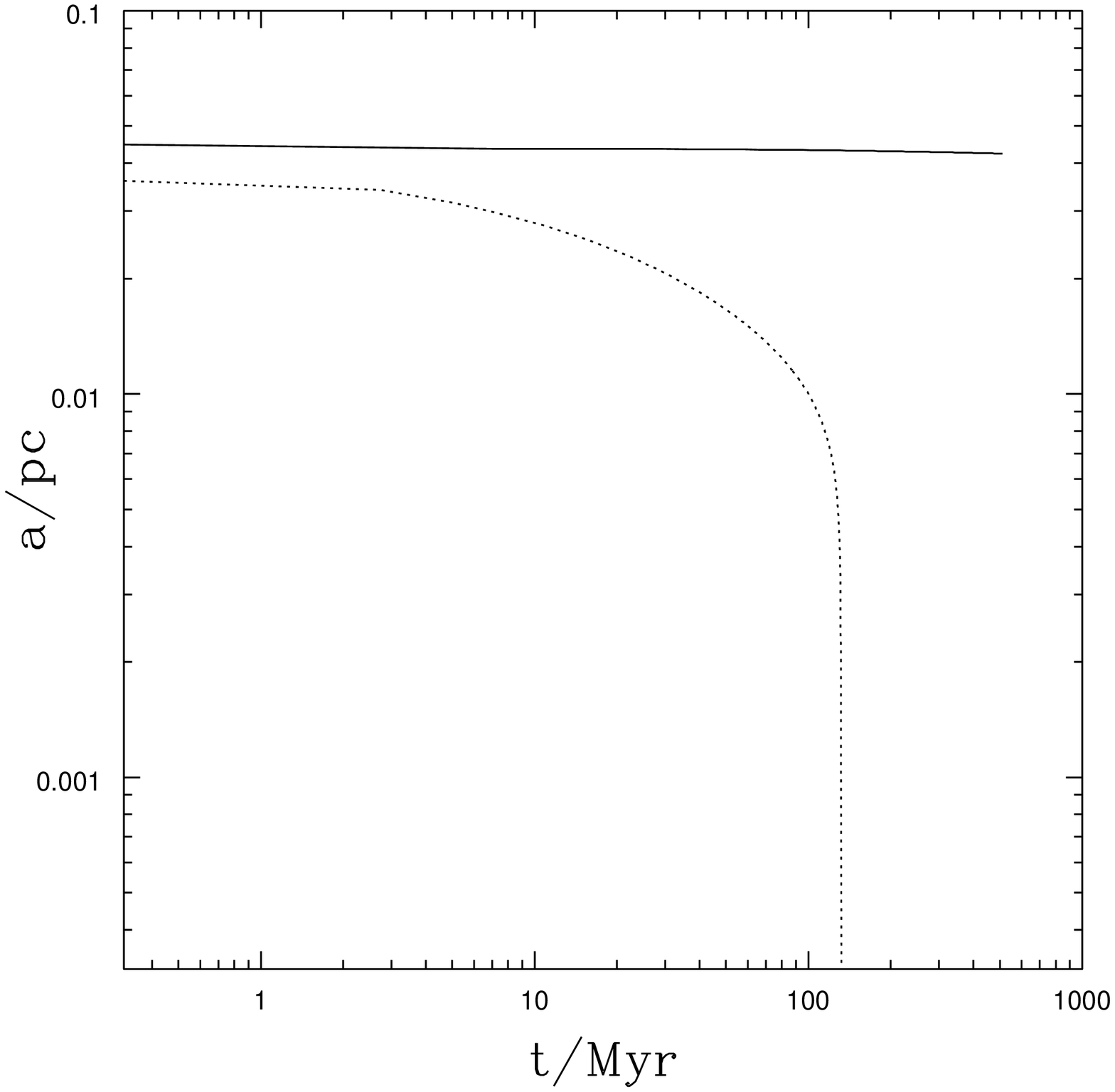}
      \caption{Same as Figure \ref{fig:a_SF}, but for the initial separation
        $a_0 = 0.05$ pc and the disc/secondary mass ratio is $M_{\rm d}/M_{\rm
          s}=1$. As before, solid curve shows the case with star formation
        included, whereas the dotted shows that where star formation ignored.
        Star formation severely depletes the disc, and the remaining disc mass
        is not large enough to induce a black hole merger within a Hubble
        time.}
	\label{fig:a_SF_pc5}
  \end{center}
\end{figure}

Fig. \ref{fig:SF} shows the evolution of the total star formation rate
within the disc. We can see that over the first $10^5$ years
the disc undergoes significant star formation. The total mass converted into
stars  is of the order of $\approx 2~10^6M_{\odot}$, that is roughly half
of the initial disc mass.  Finally, we show in Fig. \ref{fig:combine_SF} the
main properties of the disc after $\approx 6$ Myr from the beginning of the
simulation, after most of the star formation has died out but before the
secondary has migrated significantly. The top panel shows the disc surface
density, the middle panel shows the aspect ratio $H/R$ and the bottom panel
shows the value of the stability parameter $Q$. We can thus see that our
star formation prescription indeed causes the disc to hover near marginal
stability in its outer parts.

Summarising our findings for the simple energy feedback star formation model,
we note the following.  In this model, the evolution of the system does not
depend significantly on the initial disc mass (except when it is so low as to
be gravitationally stable at all radii, and would not be able to drive the
merger). Thus different initial disc masses only imply differing amounts of
star formation rate at early times, and the surface density after a few
million years is very similar in all cases.

The effect of changing the efficiency of star formation feedback
$\epsilon$ is not trivial. Naively one would expect to simply have
more star formation as one decreases its value
(cf. eq. (\ref{eq:SF})), thus leading to a smaller disc density
$\Sigma$. However, the reduced $\Sigma$ requires a smaller temperature
in order to be gravitationally stable, which in turn implies a much
smaller cooling rate and a lower star formation rate. We have run
different simulations with $\epsilon=10^{-3}$ and $\epsilon=10^{-4}$
and found only found small differences.

Finally, in Fig. \ref{fig:a_SF_pc5} the solid line shows the evolution of the
binary orbit in the presence of star formation when the initial separation is
0.05 pc, the secondary/primary mass ratio is $q=0.1$ and the initial disc mass
is equal to the secondary mass. This should be compared with the equivalent
run where star formation is not included (dotted line). Clearly,
while in principle there is initially enough gas mass in the disc to bring the
holes to coalescence, the effect of star formation depletes the disc severely
and the amount of gas left over is not enough to lead to coalescence within a
Hubble time.

\section{Discussion and conclusions}

We consider the merging of two black holes in the centre of a galaxy
driven by a gaseous disc. We make the assumption that the secondary
black hole (mass $M_{\rm s}$) arrives close to the already present
primary (mass $M_{\rm p})$ as a result of a galactic merger of some
kind \citep[cf.][]{KingPringle07}, bringing with it a certain amount
of gas (mass $M_{\rm d}$). This contrasts with the usual assumption
that the merger process manages to parachute the secondary black hole
into an already existing steady state accretion disc of infinite
extent and mass.

If the amount of gas is large ($M_{\rm d} \gg M_{\rm s}$) and does not mainly
turn into stars then the secondary black hole is simply swept into the primary
by the resulting accretion flow. Thus we consider what happens when the amount
of gas is comparable to, or smaller than, the secondary mass. We first ignore
the possibility of star formation within the accretion disc flow
(Section~\ref{sec:noSF}) in order to demonstrate the time--dependent
properties of the interactions between disc and secondary black hole. We then,
more realistically (Section~\ref{sec:withSF}), take account of formation of
stars from the disc gas when it becomes self--gravitating, and take account of
the feedback from those stars on the gas in the disc.

We have computed the properties and the SED of the disc at and around the
time of decoupling (i.e. when gravitational radiation takes over as the
dominant merging effect) in cases where the merger occurs. Our findings
differ from earlier results \citep{milosavljevic05} for three main
reasons. 

(a) The disc is not in a steady state with uniform $\dot{M}$, but
closer to a decretion disc with $\dot{M}\propto R^{-1/2}$. This
makes the SED steeper ($\nu F_{\nu} \propto \nu^{1.7}$, rather than
the steady accretion disc value of $\nu F_{\nu} \propto
\nu^{1.3}$). 

(b) The region inside the secondary orbit might not be empty, and the
small gas density can still lead to a strong burst of luminosity
during the final stages of the merger, as predicted for low
mass ratios by \citet{natarajan02}. 

(c) The enhanced dissipation resulting from the interaction between
the disc and the black hole heats the disc edges either side of the
secondary significantly. This is a new and general result, implying an
extra high photon energy contribution to the SED.

The existence of this harder emission weakens the prediction of
\citet{milosavljevic05} that the presence of a binary can be easily
inferred by a shift of the peak emission to low photon energies. More
positively, it offers a new way of inferring the presence of a binary
through a combination of this result with possible detectable periodic
variability \citep{haiman09}. Although in this paper we have assumed
that the secondary and disc orbits are exactly circular, in reality
the system may be driven to somewhat eccentric orbits. For disc
masses large enough to shrink the binary significantly (as assumed
here), dissipation in the disc is likely to keep the eccentricity low,
and so the general picture remains similar to that we have described.
However the dissipation at the inner edge of the disc will not
be uniform in azimuth or in time. This implies variability at photon
energies which one would normally associate with the innermost disc
region around a single black hole. However the typical timescale of
this disc--gap emission is not that of the innermost orbits around the
primary black hole, but instead the much longer orbital timescale of
the secondary. Thus, the detection at high photon energies of
variability on the long timescales typical of the outer disc would be
a clear signature of the presence of a black hole binary.

Evidently detailed modelling of such effects requires the use of 2D (or
even 3D) disc models, as it depends on the effectiveness of disc waves
generated by tides to carry away the excess tidal energy.

Circumbinary discs are gravitationally unstable at $R\approx 0.1$pc, just as
their cousins around single SMBH are. It is not yet clear if star formation
``catastrophe'' can be avoided due to the action of thermal or momentum
feedback from massive stars in the disc
\citep[e.g.,][]{goodman03,ThompsonEtal05}. In the case of the thermal
feedback, as shown here, gas is converted into stars too quickly compared to
the binary evolution timescales. This has a severe impact on the ability of the
disc to induce a black hole merger in a reasonable time. We find that the
rapid effect of star formation sets an upper limit to the effective mass of a
gaseous disc. Because of this, mergers within a Gigayear require initial
separations closer than $a_0 = 0.01$ pc rather than 0.05 pc, implying that gas
driven mergers of this sort are unable to solve the `final parsec' problem. To
avoid the disc becoming self-gravitating the disc must have $Q \gtrsim 1$ and
so must have a mass at most $M_{\rm d}/M_{\rm p}\approx H/R$. We also require
that the disc mass be at least comparable to the secondary mass to produce a
merger. Our modelling of star formation leads to disc aspect ratios $H/R$ of
order of a few times $10^{-3}$ and so we conclude that disc assisted mergers
only work for mass ratios $q\lesssim 0.001$ if only thermal feedback is
included. Whether such low mass secondaries would be able to produce a tight
pair in the first place is a matter of debate \citep{callegari09}.

Our treatment of star formation is very simplistic by necessity. The
inclusion of other effects neglected here, such as momentum feedback
and dynamical support from the newly born stars embedded in the disc,
might stabilse the disc at higher surface densities. Additionally, the
process of fragmentation might result in the formation of clouds and
clumps in the disc, rather than stars, whose dynamics is still
unclear. Both effects would results in a relatively more massive
gaseous disc at $\sim 0.1$ pc.  Finally, dynamical friction from the
newly born stars (not included here) might help in bringing the holes
together. While it is hard to make a precise estimate, we believe
our calculations with and without star formation (see Figures
\ref{fig:a_SF} and \ref{fig:a_SF_pc5}) bracket the range of possible
timescales for disc--induced binary mergers.

Nevertheless it seems quite likely that the `final parsec' problem is
difficult to overcome in all cases, particularly for non--extreme mass
ratios. This would imply the relative ubiquity of secondary black holes in
galactic nuclei.

\section*{Acknowledgments}

We thank the referee for a careful and constructive review of the paper.

\bibliographystyle{mn2e} 
\bibliography{lodato}

\end{document}